\documentclass[prl,aps,twocolumn,floatfix,nofootinbib]{revtex4}
\usepackage{amsmath}
\usepackage{epsfig}
\usepackage{bm}
\usepackage{slashed}
\usepackage[active]{srcltx}
\usepackage{color}
\def\tri{{{}^3{\mathrm H}}}
\def\hel{{{}^3{\mathrm {He}}}}
\def\het{{{}^3{\mathrm {He}}}}
\def\heq{{{}^4{\mathrm {He}}}}

\def\be{\begin{equation}}
\def\ee{\end{equation}}
\def\bea{\begin{eqnarray*}}
\def\eea{\end{eqnarray*}}

\def\bi{\begin{itemize}}
\def\ei{\end{itemize}}
\def\n{\phantom{1}}

\usepackage{tikz}
\usetikzlibrary{calc}
\usetikzlibrary{patterns}
\usetikzlibrary{decorations.pathmorphing}
\usetikzlibrary{decorations.markings,trees,positioning,arrows.meta}
\tikzset{
  box/.style   = { rectangle, rounded corners, draw, align=left},
  valve/.style = {-{Triangle[fill=white,width=0.5cm,
      length=0.5cm]Triangle[fill=white,width=0.5cm,length=0.5cm,reversed]}},
}
\tikzset{
  double arrow/.style args={#1 colored by #2 and #3}{
    -stealth,line width=#1,#2, % first arrow
    postaction={draw,-stealth,#3,line width=(#1)/1.8,
                shorten <=(#1)/4,shorten >=2*(#1)/4}, % second arrow
  }
}

\begin{document}
\title{The $\het(\vec {\mathrm n},{\mathrm p})\tri$ parity-conserving asymmetry}

\author{
M. Viviani$^{1}$\footnote{Corresponding author, email: michele.viviani@pi.infn.it}, S. Bae\ss{}ler$^{2,3}$,
L. Barr\'on-Palos$^4$, N. Birge$^{5,6}$, J.D. Bowman$^3$, J. Calarco$^7$,
V. Cianciolo$^3$, C.E. Coppola$^{5}$, C.B. Crawford$^8$, G. Dodson$^9$, N. Fomin$^5$, I. Garishvili$^{3,5}$,
M.T. Gericke$^{10}$, L. Girlanda$^{11,12}$, G.L. Greene$^{5,3}$,
G.M. Hale$^6$, J. Hamblen$^{13}$, C. Hayes$^{5,14}$, E. B. Iverson$^{3}$, M.L. Kabir$^{8,15}$, A. Kievsky$^{1}$,
L.E. Marcucci$^{16,1}$, M. McCrea$^{10,17}$ E. Plemons$^{5}$,
A. Ram\'irez-Morales$^4$, P.E. Mueller$^3$, I. Novikov$^{18}$, S.I. Penttila$^3$, E.M. Scott$^{5,19}$, J. Watts$^{13}$,
and C. Wickersham$^{13}$ 
}
\affiliation{
$^1$ Istituto Nazionale di Fisica Nucleare, Sezione di Pisa, I-56127 Pisa, Italy\\
$^2$ University of Virginia, Charlottesville, Virginia 22904, USA\\
$^3$ Oak Ridge National Laboratory, Oak Ridge, Tennessee 37830, USA\\
$^4$ Instituto de F\'{i}sica, Universidad Nacional Aut\'{o}noma de M\'{e}xico, Apartado Postal 20-364, 01000, M\'{e}xico\\
$^5$ University of Tennessee Knoxville, Knoxville, Tennessee 37996, USA\\
$^6$ Los Alamos National Laboratory, Los Alamos, New Mexico 87545, USA \\
$^7$ University of New Hampshire, Durham, New Hampshire 03824, USA\\
$^8$ University of Kentucky, Lexington, Kentucky 40506, USA\\
$^9$ Massachusetts Institute of Technology, Cambridge, Massachusetts 02139, USA\\
$^{10}$ University of Manitoba, Manitoba  R3T 2N2, Canada\\  
$^{11}$ Department of Mathematics and Physics, University of Salento, I-73100 Lecce, Italy \\
$^{12}$ Istituto Nazionale di Fisica Nucleare, Sezione di Lecce,  I-73100 Lecce, Italy \\  
$^{13}$ University of Tennessee Chattanooga, Chattanooga, Tennessee 37403, USA \\
$^{14}$ Indiana University, Bloomington, Indiana 47405, USA\\
$^{15}$ Brookhaven National Laboratory, Upton, NY 11973, USA\\  
$^{16}$ Department of Physics ``E. Fermi'', University of Pisa, I-56127, Pisa, Italy\\
$^{17}$ University of Winnipeg, Winnipeg, Manitoba R3T 2N2, Canada \\
  $^{18}$ Western Kentucky University, Bowling Green, Kentucky, 42101, USA\\
  $^{19}$ Centre College, Danville, KY 40422, USA
}

\begin{abstract}
  Recently, the n$^3$He collaboration reported a measurement of the parity-violating (PV) proton directional asymmetry $A_{\mathrm {PV}} = (1.55\pm 0.97~\mathrm {(stat)} \pm 0.24~\mathrm {(sys)})\times 10^{-8}$ in the capture reaction of $\het(\vec {n},{\mathrm p})\tri$ at meV incident neutron energies. The result
  increased the limited inventory of precisely measured and calculable
  PV observables in few-body systems required to further understand the structure of hadronic weak interaction. In this letter, we report the experimental and theoretical investigation of a parity conserving (PC) asymmetry $A_{\mathrm {PC}}$ in the same reaction (the first ever measured PC observable at meV neutron energies).  
  As a result of S- and P-wave mixing in the reaction, the $A_{\mathrm {PC}}$ is inversely proportional to the neutron wavelength $\lambda$. The experimental value is
  $ (\lambda\times A_{\mathrm {PC}})\equiv\beta = (-1.97 \pm 0.28~\mathrm {(stat)} \pm 0.12~\mathrm {(sys)}) \times 10^{-6}\;\mathrm{\AA}$.
  We present results for a theoretical analysis of this reaction by solving the four-body scattering problem within the hyperspherical
  harmonic method.  We find that in the $\het(\vec {n},{\mathrm p})\tri$ reaction,  $A_{\mathrm {PC}}$ depends critically
  on the energy and width of the close $0^-$ resonant state of $\heq$, resulting in a large sensitivity to the spin-orbit components
  of the nucleon-nucleon force and even to the three-nucleon force.
  The analysis of the accurately measured $A_{\mathrm {PC}}$ and $A_{\mathrm {PV}}$ using the same few-body theoretical models 
  %methods%
  gives essential information needed to interpret the PV asymmetry in the 
  $\het(\vec {n}, {\mathrm p})\tri$ reaction.
  %
  %Consequently, the study of this observable can give {\Red{important}}
  %insight into the nuclear dynamics
  %, necessary, in particular, to develop models and extract PV observables from measurements on
  %nuclear systems.
  %
\end{abstract}

%\date{\today}

\maketitle

%\section{Introduction}
%\label{sec:intro}

\clearpage

{\bf Introduction.} The study of polarization observables in nuclear reactions is
an important tool - in some cases the only tool - to improve our understanding on issues ranging from fundamental symmetries to still ambiguous observations in the strong nuclear interaction. In this letter we present results from an investigation of the reaction of $\het(\vec {n},{\mathrm p})\tri$ using transverse polarized cold neutrons (i.e. of meV energies) at the Spallation Neutron Source (SNS) of the Oak Ridge National Laboratory (ORNL). In a previous paper, we reported the parity-violating (PV) asymmetry $A_{\mathrm {PV}} = (1.55\pm 0.97~\mathrm {(stat)} \pm 0.24~\mathrm {(sys)})\times 10^{-8}$~\cite{n3he20}. Here, we present the experimental and theoretical investigation of a parity-conserving (PC) asymmetry in the same reaction.
%
%We also report theoretically investigate of  its dependence on different interaction models.
%using transverse polarized cold
%neutrons at the Spallation Neutron Source (SNS) of the Oak Ridge National Laboratory (ORNL), %for details see~\cite{n3he20}.
%
In general, the cross section for $\het(\vec {n},{\mathrm p})\tri$ can be written as
\be
   {\mathrm d\sigma\over \mathrm d\Omega}=\left({\mathrm d\sigma\over \mathrm d\Omega}\right)_\mathrm u
   \Bigl(1+A_{\mathrm {PV}}\hat s_\mathrm n\cdot \hat k_\mathrm p + A_{\mathrm {PC}} (\hat s_\mathrm n\times \hat k_\mathrm n) \cdot \hat k_\mathrm p\Bigr) \ , \label{eq:def}
\ee
where $(\mathrm d\sigma/\mathrm d\Omega)_\mathrm u$ is the unpolarized cross section and $\hat s_\mathrm n$, $\hat k_\mathrm p$, and $\hat k_\mathrm n$ denote unit vectors
specifying the directions of the neutron polarization, the outgoing proton momentum, and the incoming neutron beam, respectively.
The $A_{\mathrm {PC}}$ is measured by detecting emitted protons with their momenta in the plane defined by $\hat s_\mathrm n \times \hat k_\mathrm n$ and $\hat k_\mathrm n$, and the $A_{\mathrm {PV}}$ in the plane $\hat s_\mathrm n$ and $\hat k_\mathrm n$. The asymmetry is then deduced from detector yields with opposite neutron spin direction.

At a fundamental level, PV observables in nuclei are a consequence of the hadronic weak interaction (HWI) between quarks, which explains their very small values, see, for example, Ref.~\cite{Vries20} for a recent review. Interest in measurements of $A_{\mathrm {PV}}$ in nucleon-nucleon (NN) systems (or in light nuclei) at low incoming neutron energies is therefore motivated by the effort to find additional insight to the structure of the HWI, the least-known part of the weak interaction~\cite{MP06,HH13,SS13,DDH,KS93}. Using the same beam and the $A_{\mathrm {PV}}$ experimental setup, the collaboration was able to measure $A_{\mathrm{PC}}$, just by changing the plane of the detection of the emitted protons~\cite{n3he20}. This is an important feature when systematic uncertainties are considered.
As shown below, the observable is directly sensitive to the strong and electromagnetic components of the nuclear interaction, and contributions from the HWI can be safely neglected because they are estimated to be a few orders of magnitude smaller. Because $A_{\mathrm {PC}}$ is a consequence of the interference between S-- and P--waves of the incoming neutron at meV energies, 
$A_{\mathrm {PC}}\propto 1/\lambda$, where $\lambda$ is the neutron wavelength.  
%Wave vector of a cold neutron with $\lambda = 5\times 10^5$ fm is $k= 1.3\times %10^{-5}$ fm${}^{-1}$ and using the ${}^3$He radius of $R\approx1.97$ fm, we obtain %$(kR)\sim 2\times 10^{-5}$. The small value predicts that the P-wave contributions of %the incident neutrons to the reaction can be strongly suppressed.
%Since $\lambda\sim 10^{5}$~fm is quite large on the scale of nuclear physics (whose typical ranges are of %the order of a few fm), $A_{\mathrm {PC}}$ is expected to be small. 
More specifically, the scale of the $A_{\mathrm {PC}}$ will be proportional to $(kR)$, where $k=2\pi/\lambda$ and $R$ a characteristic length for this reaction. For $\lambda=5 \times 10^5$ fm, the neutron wave vector is $k \sim 1.3 \times 10^{-5}$ fm$^{-1}$ and using $R \sim 1.97$~fm for the $^3$He radius, we obtain $(kR) \sim 2 \times 10^{-5}$. This is the expected order of magnitude of $A_{\mathrm {PC}}$.
%The small value predicts that contributions of incident P--waves to the reaction are strongly suppressed, see theory section for details.

For the unpolarized $\het(n,\mathrm p)\tri$ reaction, measurements of the total cross section and the differential cross section exist at very low energies. No data for PC polarization observables were reported for this reaction, only for the mirror reaction $\tri(p,\mathrm n)\het$~\cite{exfor,viviani20}. These experiments were performed at proton incident energies corresponding to neutron energies of 300~keV or greater. Therefore, the measurement that we report here, is the first ever measurement of a PC $\het(n,\mathrm p)\tri$ polarization observable at meV incident neutron energies.

Since the four-nucleon scattering problem can be routinely solved, $A_{\mathrm {PC}}$ can be accurately calculated starting from a given model of the strong and electromagnetic interactions~\cite{DF07,DF17,Lazaus09,LC20,PHH01,Sofia10,Aoyama11,bm16}. This observable is usually very sensitive to the nuclear interaction, in particular, to spin-orbit components of the NN interaction and the three-nucleon (3N) force. This can be readily understood, since $A_{\mathrm {PC}}$ is a consequence of the interaction in P-waves. It is worth mentioning that in the physics of few-nucleon systems, we still have various discrepancies between theory and experiment, such as the famous ``$A_{\mathrm y}$-puzzle'' in ${{p-d}}$, ${{n-d}}$, and ${ {p-\het}}$ scattering~\cite{KH86,WGC88,Kie96,Fon99,DF07,Vea13}. In the present case, this sensitivity could be amplified since the process under study is at an energy rather close to the energy of the second excited state of $\heq$, that has quantum numbers $J^\pi=0^-$~\cite{Tilley92}. Therefore, the study of this PC observable could be an extraordinary opportunity to study this poorly known resonance.

The energy spectrum of $\heq$ is, in fact, an important testing ground for understanding nuclear dynamics.
Energies and widths of the various resonances have been determined in R-matrix analyses~\cite{Tilley92}.
As a matter of fact, all excited states are resonances; however, their precise energies and widths contain critical information. Their calculation using different Hamiltonians gives slightly different results~\cite{viviani20}.
The first excited state, a $0^+$ resonance, has been vigorously investigated both theoretically and
experimentally. Without the Coulomb interaction, it would be a true bound state, with an energy
well in agreement with that predicted in the framework of $A=4$ Efimov physics~\cite{Kie21}. This state has been studied experimentally by means of electron scattering, see Ref.~\cite{Kegel21} for a recent analysis. Theoretical studies
have found that the position and width of this resonance are critically dependent on the interaction~\cite{Hiyama04,Bacca13,Bacca14,Michel23,M23}. The next excited state, the $0^-$ resonance, is just above the threshold of ${n}-\het$ dissociation, and has a similar width as the $0^+$ resonance. Its existence is due to the interaction in P-waves.

{\bf Measurement and analysis of $A_{\mathrm {PC}}$.}~The primary goal of the n${}^3$He experiment was a precision measurement of $A_{\mathrm {PV}}$; however, $A_{\mathrm {PC}}$ was also measured to support the theoretical analysis of the $n-^3$He reaction and because it is a significant correction to $A_{\mathrm {PV}}$, see Table~1 in Ref.~\cite{n3he20}. To keep sources of systematic uncertainties in the $A_{\mathrm {PC}}$ and  $A_{\mathrm {PV}}$ measurements comparable, the two measurements were designed 
so that the only change from the $A_{\mathrm {PV}}$ setup was a rotation of the target/detector chamber around the beam axis by $90$~degrees~\cite{n3he20}, to the most sensitive detector orientation for $A_{\mathrm {PC}}$. We reported the uncorrected $A_{\mathrm {PC}}^{\mathrm {unc}} = (-41\pm5.6 (\mathrm {stat}))\times10^{-8}$ in Ref.~\cite{n3he20}, extracted from the data with the same algorithm that was used for the $A_{\mathrm {PV}}$ data, where the asymmetry was calculated from integrating detector yields over a neutron wavelength range,~{3.4--6.3~\AA}~\cite{n3he20}. This averaging approach was possible in the case of the  PV asymmetry, since it does not depend on incident neutron energy. However, $A_{\mathrm {PC}}$ has the $1/\lambda$ dependence that is incorporated to analysis of the $A_{\mathrm {PC}}$ data in this letter.

Corrections and systematic uncertainties for $A_{\mathrm {PC}}^{\mathrm {unc}}$ are mainly due to the uncertainty in the chamber orientation. 
The first two corrections are produced by a twist of the signal wire frame stack in the detector~\cite{n3he20} and then the alignment uncertainty of the spin holding magnetic field with the detector. These two alignment uncertainties mix the small PV value into the significantly larger PC value, and are therefore expected to be small. 
We get the frame twist and the field alignment correction to $A_{\mathrm {PC}}^{\mathrm {unc}}$ to be $-0.09 \pm 0.00$~ppb and $0.00 \pm 0.04$~ppb, respectively~\cite{n3he20}. 

\begin{figure}
\includegraphics[trim={0 3.2 0 0},clip,width=0.48\textwidth]{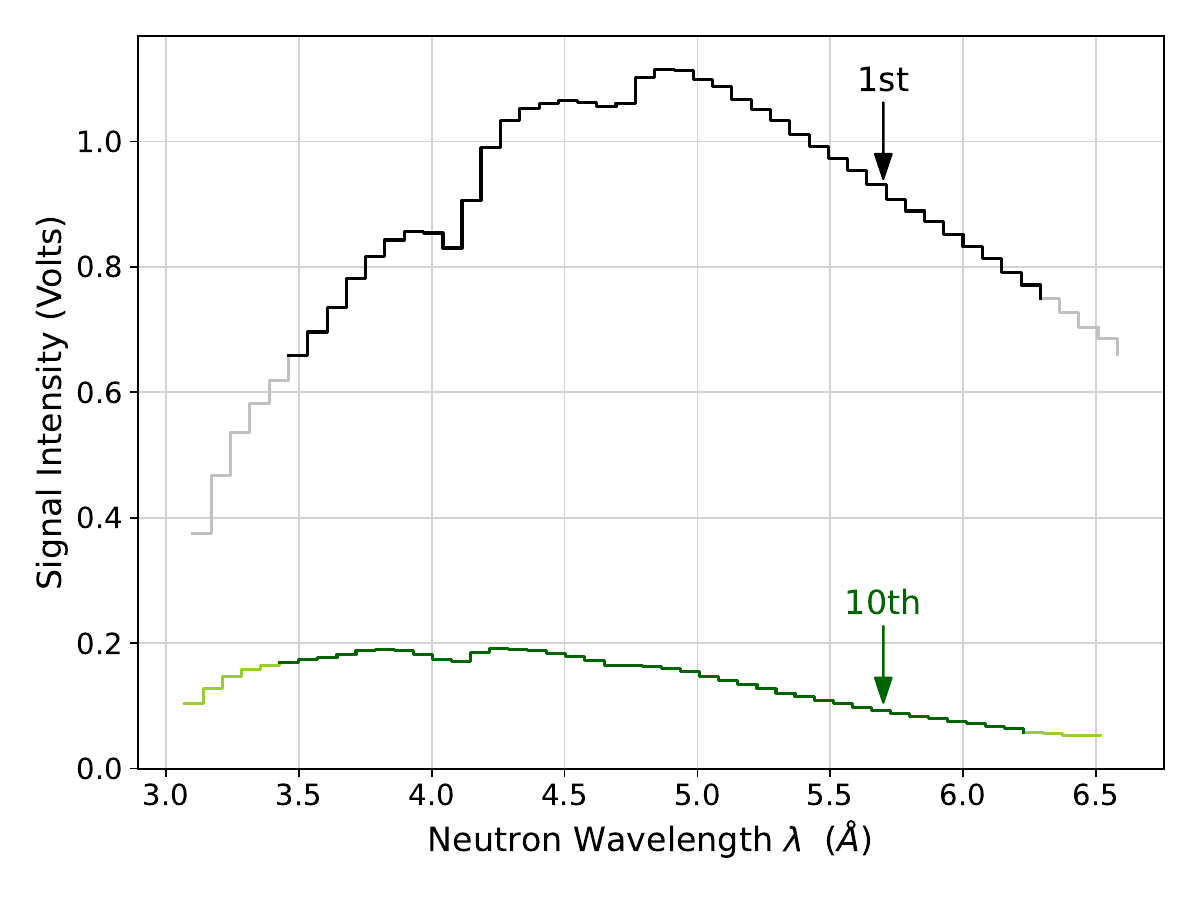}
\caption{ The chamber has 16 signal wire planes. The measured outputs of the central signal wires of the
$1^{\mathrm{st}}$ and the $10^{\mathrm{th}}$ wire plane have selected to indicate how the incoming neutron intensity and peak shape changes as the beam passes through the chamber. 
The signals were measured as a function of $TOF$ but are plotted as a function of the neutron wavelength.
The DAQ samples the wires at $20~\mu$s intervals and then sums them into 49, 0.32~ms wide time bins.
Due to $17$~cm difference in distance between 
%the $1^{\mathrm {st}}$ and the $10^{\mathrm {th}}$ 
the two wire planes from the moderator, the neutron wavelengths on the $10^{\mathrm {th}}$ are little shorter than on
the $1^{\mathrm {st}}$ wire, consequently, there is a half time bin shift between the wavelength plots.
The signals represent collected charges after the n$^3$He capture and the
emitted proton and triton depositing their kinetic energy to $^3$He gas at 43.6 kPa.
The difference between the two spectra is mainly due to a large neutron energy
dependent capture cross section on $^3$He that reduces exponentially the intensity
of the beam traveling down the chamber. For determination
of $\beta$, see text for details, the integration is done over the same time bin range,
meaning that each wire plane has a unique $\lambda_{\mathrm {min}}$ and $\lambda_{\mathrm {max}}$ for integration. The
integration ranges of the two signals are indicated with thicker lines. The two
small dips at $4.1~{\mathrm {\AA}}$ and $4.7~{\mathrm {\AA}}$ are residuals from neutron transmission Bragg-edges on Aluminum beamline windows.}

	\label{fig:PulseShape}
\end{figure}

The origin of the third additive correction is the spin-orbit component of the electromagnetic neutron-${}^3$He atom interaction due to the Mott-Schwinger (MS) mechanism~\cite{Gericke08}. In fact, the probability that a neutron undergoes an elastic scattering on ${}^3$He atoms in the target before being captured by another helium nucleus, is very small, $\sim 10^{-4}$, but, in any case, finite. This elastic scattering produces, via the MS mechanism, a small left-right asymmetry $A_{\mathrm {MS}}$ that is then ``conserved'' in the subsequent capture, which is essentially isotropic. The sequence (elastic scattering + capture) produces at the end a false asymmetry $\delta A_{\mathrm {MS}}$, which has to be subtracted from  $A_{\mathrm {PC}}^{\mathrm {unc}}$ in order to obtain the PC asymmetry due solely to the capture process. We calculated $A_{\mathrm{MS}}$ following the approach described in Ref.~\cite{Gericke08}. Since it depends on the neutron wavelength $\lambda$, we have to average over the measured wavelength range at each wire plane $\mathrm j$ (the geometry of the detector chamber is briefly discussed in the caption of Fig.~\ref{fig:PulseShape}), i.e.
\begin{equation}
  \bar{A}^\mathrm j_{\mathrm{MS}}= \int_{\lambda^\mathrm j_{\mathrm {min}}}^{\lambda^\mathrm j_{\mathrm {max}}} A^\mathrm j_{\mathrm{MS}}(\lambda) P^\mathrm j(\lambda){\mathrm{d}}\lambda\ ,\label{eq:avems}
  \end{equation}
where $P^\mathrm j(\lambda)$ is the yield density distribution of the wire plane. 
Fig.~\ref{fig:PulseShape} shows yields from two central signal wires of the $1^{\mathrm {st}}$ and the $10^{\mathrm {th}}$ plane. The target chamber has total of 16 planes 19~mm apart. $\lambda^\mathrm j_{\mathrm {min (max)}}$ are the integration limits, unique in each wire plane due to the increasing distance of the planes from the moderator.
The significant variation in the detector yield between the planes is due mainly to the large wavelength dependent n-$^3$He capture cross section \cite{n3He cross sections92,Als-Nielsen64} that exponentially decreases the beam intensity as it passes through the $\hel$ gas at a pressure of $43.6$~kPa. 
The overall MS asymmetry $\overline A_{\mathrm{MS}}$ for the detector is obtained by calculating the average given in Eq.~\eqref{eq:avems} for each of the 16 wire planes and then combining those using the integrated yield in the respective planes as the weighting factor.
The final calculated value is $\overline A_{\mathrm {MS}}=(-4.8 \pm 0.2) \times 10^{-5}$, that results in a correction $\overline{\delta A}_{\mathrm{MS}} = (-1.36 \pm 0.07) \times 10^{-8}$ to $A_{\mathrm{PC}}$. With the above corrections and those from Table ~1 of Ref.~\cite{n3he20}, we obtain the final value for the measured PC asymmetry 
\begin{equation}
  A_{\mathrm{PC}}= (-42.4\pm 6.0 (\mathrm {stat})\pm 0.23(\mathrm {sys}))\times 10^{-8}\,,\label{eq:Apa}
\end{equation} 
where the error in the calculated value of $\overline{\delta A}_{\mathrm{MS}}$ has been folded into the systematic uncertainty. 

%%before the MS correction (-43.9 +- 6 +- 0.2) x 10-8

The incoming neutrons contributing to the PC asymmetry reported in Eq.~(\ref{eq:Apa}), have a range of wavelengths. Since $A_{\mathrm {PC}}~\propto~1/\lambda$, we can write explicitly
\begin{align}
A_{\mathrm {PC}}(\mu) = \beta\mu, 
\label{eq:betaRelation}
\end{align}
where $\beta$ is a constant that we want to extract from $A_{\mathrm {PC}}$  and $\mu= 1/\lambda$. The mean inverse wavelength over the integrated wave forms in each wire plane can be calculated as discussed previously in context of Eq.~(2) and in the Appendix. For the overall  $\overline\mu$ we obtain
\begin{equation}
  \overline\mu=\left\langle{1\over\lambda}\right\rangle=0.2151 \pm 0.0015 \;\mathrm{\AA}^{-1}\,,\label{eq:mua}
  \end{equation}
where the uncertainty is statistical, and is determined by the averaging process. 
With Eqs.~(3)--(5), we get
\begin{equation}
\beta= (-1.97 \pm 0.28 (\mathrm{stat}) \pm 0.02 (\mathrm{sys})) \times 10^{-6}\; \mathrm{\AA}\,.\label{eq:betaa}
\end{equation}
However, to finalize systematic uncertainty in $\beta$, we still need to work out the full uncertainties in $\overline\mu$.

In this pulsed neutron source, the wavelength of neutrons arriving at a detector location is calculated  by the measured time of flight ($TOF$) over the known path length to that detector. In our case, the path length from the moderator emission surface to the $1^{\mathrm {st}}$ wire plane has been measured with relative accuracy of $0.2$ \%. The $TOF$ measurement is started by a clock signal corresponding to the arrival of the proton beam bunch on the mercury target to within 1~${\mathrm {\mu}}$s, the neutron emission time - a small correction to $TOF$ due to time from start of $TOF$ to moment when the neutron exits the moderator~\cite{{emissiontime}}. After a selected delay (13.500~ms), the DAQ system digitizes the signals in 49 time bins of 0.32~ms width. Fig.~1 shows typical yields, measured in the $1^{\mathrm {st}}$ and $10^{\mathrm {th}}$ wire planes as a function of $TOF$ but presented as a function of wavelength. This presentation required that the plots had to be shifted little respect to the x-axis, since the two planes are at different distances from the moderator. The $TOF$ calculation includes, as its most significant uncertainty, the uncertainty of the charge collection time, the time that is required to collect all the charges from the ionization by the emitted proton from the $n-{}^3$He capture. Results of Garfield++ simulations indicate that the collection time can be up to $1.5$~ms long~\cite{Mark2016}. This is the dominating uncertainty in $\overline\mu$ and is carried over to the systematic uncertainty in $\beta$. A more detailed discussion of the determination of the uncertainty is contained in the Appendix. Our final result for $\beta$ is 
\begin{align}
	\beta = (-1.97\pm 0.28~\mathrm {(stat)}\pm 0.12~\mathrm {(sys)})\times 10^{-6}\;\mathrm{\AA}\,.
 \end{align}

{\bf Theoretical analysis.} In the present paper the four-nucleon (4N) scattering problem is solved using the
hyperspherical harmonics (HH) method~\cite{viviani20,Mea20}. Here, we will 
give only some details of the procedure.
First, a particular clusterization $A+B$ of the four-nucleon system in the asymptotic region is denoted with the index $\gamma$.
More specifically, $\gamma=1$ ($2$) corresponds to $p-\tri$ ($n-\het$) clusterization.
Consider a scattering state with total angular momentum quantum number
$JJ_z$, and parity $\pi$. The asymptotic part of the wave function describing
the incoming clusters $\gamma$ with relative orbital angular momentum $L$ and total spin $S$ (note that $\pi\equiv(-)^L$) is generally written as a sum of a (distorted) plane wave plus outgoing spherical waves in all possible channels $\gamma',L',S'$. The weights of the outgoing waves are the T-matrix elements (TMEs)  ${}^JT^{\gamma,\gamma'}_{LS,L'S'}$. It is well known that ${}^JT^{2,1}_{LS,L'S'}\sim q_2^L$ for $q_2\rightarrow0$~\cite{viviani20}, where $q_2$ is the relative $n-\het$ momentum. To complete the calculation, the inner part of the wave function (where the nucleons are close between themselves) is described by an expansion over the HH basis (for more details, see Ref.~\cite{viviani20}).

In terms of the TMEs, the transverse asymmetry can be written as
 \begin{eqnarray}
  \lefteqn{ A_{\mathrm{PC}}\sigma_0 =\qquad\qquad}&&  \nonumber \\
  && 3\sqrt{2} \Im\left[(^1T^{2,1}_{11,10})^*\; {}^0T^{2,1}_{00,00}\right] +2\Im\left[({}^0T^{2,1}_{11,11})^* \; {}^1T^{2,1}_{01,01}\right] 
   \nonumber \\
   +&& 3\sqrt{2} \Im\left[(^1T^{2,1}_{10,11})^* \; {}^1T^{2,1}_{01,01}\right]  +3\sqrt{2} \Im\left[(^1T^{2,1}_{11,11})^* \; {}^1T^{2,1}_{01,01}\right] \nonumber\\
   -&& 5 \Im\left[(^2T^{2,1}_{11,11})^*\; {}^1T^{2,1}_{01,01}\right]\ , \label{eq:att}\\
   \sigma_0 &\!\!=\!\!& |^0T^{2,1}_{00,00}|^2 + 3 |{}^1T^{2,1}_{01,01}|^2    \ ,\label{eq:sigma0}
 \end{eqnarray}
where in the expression of $\sigma_0$ we have retained only the contributions of S-wave TMEs (which are $\sim q_2^0$), and in $A_{\mathrm {PC}}$ the contribution of S-- and P--waves.  From the $q_2$ behavior discussed above, we
see that $A_{\mathrm {PC}}\sim q_2$. Note that $q_2=(3/4) 2\pi/\lambda$, hence $A_{\mathrm {PC}}\sim 1/\lambda$ as discussed previously.
 
The $A_{\mathrm {PC}}$ observable has been obtained using the NN interaction derived 
by Entem and Machleidt at next-to-next-to-next-leading order (N3LO) in chiral effective field theory (EFT)~\cite{EM03,ME11}, corresponding to 
two different cutoff values ($\Lambda=500$ MeV and
$\Lambda=600$ MeV). These NN interactions are labeled,
respectively, N3LO500 and N3LO600.  In this way, we can explore the
dependence on the cutoff value $\Lambda$ of the 4N observables. 
The 3N force considered here has been derived at next-to-next-leading order (N2LO) in
Ref.~\cite{Eea02} (the 3N force contributions at N3LO and beyond are still under
construction but we plan to include them in future 4N calculations).
With the N3LO500 (N3LO600) NN interaction, we have considered the 3N
N2LO force, in the local coordinate space version~\cite{N07}, labeled N2LO500 (N2LO600) with the parameters $c_D$ and $c_E$
fixed to reproduce the 3N binding energy and the experimental Gamow-Teller (GT) matrix element
in the tritium beta decay. 
The values of these parameters, recently redetermined in Ref.~\cite{Mea18b}, are
$(c_D,c_E)=(+0.945,-0.0410)$ for the N2LO500 force and
$(c_D,c_E)=(+1.145,-0.6095$) for the N2LO600 force. 

In order to explore the dependence on the
parameters $c_D$ and $c_E$, we use also another 3N N2LO force labeled N2LO500*.
In this case, the interaction reproduces the 3N binding energies but not the tritium GT matrix element. The corresponding values of the parameters are chosen arbitrarily to be $(c_D,c_E)=(-0.12,-0.196)$.

We also report results obtained using the Norfolk NN interactions, also derived within chiral EFT, but using as degrees of freedom nucleons, pions and $\Delta$s~\cite{Piarulli16,Piarulli17}. In this case, the potentials are regularized in coordinate space. We use the so-called  NVIa and NVIb NN interactions, regularized with cutoff $R_L=1.2$ and $1.0$ fm, respectively. They are augmented by 3N N2LO interactions, with $(c_D,c_E)=(-0.635,-0.090)$ for the NVIa force and $(c_D,c_E)=(-4.71,+0.55)$ for the NVIb force (see Table IV of Ref.~\cite{Baroni18}). 

In all cases, the electromagnetic force between the nucleons has been approximated by the point-Coulomb potential. Effects due to other terms, as the magnetic dipole interaction, the vacuum polarization term, etc. are thought to be very small in capture observables.

For each interaction, the convergence of the calculated TMEs has been checked by increasing the size of the HH basis. The only problematic quantity to be calculated has been found to be the $0^-$ TME, due to the difficulty of constructing the 4N wave function close to the $0^-$ resonance. 

In order to explore the properties of this resonance, we have performed a series of calculations for different neutron energies. 
From the TMEs we can extract also the resonance position $E_{\mathrm R}$ and width $\Gamma$, using the procedure described
in Ref.~\cite{viviani20}. Results are reported in columns 2 and 3 of Table~\ref{tab:res}. 
The values are generally smaller than those extracted from the experiment using an R-matrix analysis~\cite{Tilley92}
(actually, the two methods do not necessarily give the same result). 
The $E_{\mathrm R}$ value in correspondence with the N3LO500/N2LO500 interaction is found to be smaller than for the other cases. Also the width $\Gamma$ is found to be narrower.

 \begin{table}[t]
\caption{\label{tab:res}
  Results of the calculations performed using the HH method. In columns 2 and 3,
  we report the position $E_R$ and width $\Gamma$ of the $\heq$ $0^-$ resonance obtained as discussed in Ref.~\protect\cite{viviani20}.
  Note that $E_{\mathrm R}$ is calculated
  starting from the $n-\het$ threshold (nearly $20.58$ MeV above the $\heq$ ground state energy).
  In column 4, we report the values of
  ${}^0t^{2,1}_{11,11}=\lim_{q_2\rightarrow0} | {}^0T^{2,1}_{11,11}|/q_2$
  and in column 5 the quantity $\beta$ as defined in Eq.~(\ref{eq:betaRelation})
  (we remember that this quantity at low energies is independent of $q_2$). In the last row, we report the experimental values for $E_{\mathrm R}$, 
  $\Gamma$~\protect\cite{Tilley92},  and $\beta$, the latter quantity obtained as discussed in the main text.
 }
\begin{center}
\begin{tabular}{lcccr}
\hline 
\hline 
Interaction & $E_{\mathrm R}$  & $\Gamma$ & ${}^0t^{2,1}_{11,11}$ & $\beta\times 10^6$  \\
     & [MeV] &  [MeV] & [fm] & [$\mathrm{\AA}$] \\
\hline
N3LO500 & $0.16$ & $0.41$ & $20.7$ & $\n-4.83$\\
N3LO600 & $0.24$ & $0.51$ & $16.9$ & $\n-2.68$\\
NVIa    & $0.31$ & $0.53$ & $17.3$ & $\n-2.61$ \\
NVIb    & $0.30$ & $0.54$ & $13.0$ & $\n-0.43$\\
\hline
N3LO500/N2LO500   & $0.06$ & $0.26$ & $30.1$ & $-10.04$\\
N3LO500/N2LO500*  & $0.14$ & $0.41$ & $18.5$ & $\n-2.68$\\
N3LO600/N2LO600   & $0.09$ & $0.30$ & $25.9$ & $\n-5.28$\\
NVIa/3N           & $0.04$ & $0.36$ & $35.4$ & $-12.17$\\
NVIb/3N           & $0.12$ & $0.40$ & $23.9$ & $\n-5.15$\\
\hline
Experimental  & $0.44$& $0.84$& & $-1.97$ \\
  & & & & $\pm 0.28$ ($\mathrm {stat}$) \\
  & & & & $\pm 0.12$ ($\mathrm {sys}$)\\
\hline
\end{tabular}
\end{center}
\end{table}
We note that $A_{\mathrm{PC}}$ is very sensitive to the value of ${}^0t^{2,1}_{11,11}=\lim_{q_2\rightarrow0} |{}^0T^{2,1}_{11,11}|/q_2$, reported in the $4^{\mathrm {th}}$ column of the table. Note the large variation with the different interactions. Tiny differences of the position of the $0^-$ resonance reflect in large changes in ${}^0t^{2,1}_{11,11}$.  

Finally, in column 5 we report the calculated values of $\beta=A_{\mathrm{PC}}\times\lambda$ for the various interactions using Eq.~(\ref{eq:att}). The calculations have been performed at $q_2=0$ (namely, at $\lambda\rightarrow\infty$), but in any case, for these energies $\beta$ is independent of $q_2$.
The large differences between the results for $\beta$ reflect the fact that 
the various terms in Eq.~(\ref{eq:att}) tend to cancel each other, in particular the first two, which are
the largest ones. Furthermore, note that without the contribution of the ${}^0T^{2,1}_{11,11}$ term, $A_{\mathrm {PC}}$ would be positive, at variance with what found 
 experimentally. The $A_{\mathrm {PC}}$ with N3LO500/N2LO500 is found to be too large (in absolute value), due to the
 large value of ${}^0T^{2,1}_{11,11}/q_2$. This is probably a consequence of the fact that for this interaction the $0^-$ resonance
 is very close to the $n-\het$ threshold. 

 The interaction N3LO500/N2LO500* differs from N3LO500/N2LO500 just for $c_D$, $c_E$ values. In this case $A_{\mathrm {PC}}$ is found to be in good agreement with the experimental value. This result shows the sensitivity of this observable to the details of the 3N force. The difference between the N3LO500/N2LO500 and N3LO600/N2LO600 results
 shows the sensitivity of this observable to the cutoff values and consequently
 to the different treament of the short-range physics.

%\section{Conclusions}
%\label{sec:conc}

{\bf Conclusions.} The first accurate measurement of a parity-conserving proton directional asymmetry at reaction of $\het(\vec {n},{\mathrm p})\tri$ at meV incident neutron energies resulted in $A_{\mathrm{PC}}= (-42.4\pm 6.0 (\mathrm {stat})\pm 0.23(\mathrm {sys}))\times 10^{-8}$. Since the $A_{\mathrm {PC}}$ depends on $1/\lambda$, where $\lambda$ is neutron wavelength, we can remove the $\lambda$ dependence and obtain a constant
$\beta=(\lambda\times A_{\mathrm {PC}}) = (-1.97 \pm 0.28~\mathrm {(stat)} \pm 0.12~\mathrm {(sys)}) \times 10^{-6}\;\mathrm{\AA}$. In this reaction, $A_{\mathrm {PC}}$ comes out from the S– and P–wave interference induced mainly by the strong nuclear interaction.
%A consequence of the low incident neutron energy is that the value of the $A_{\mathrm{PC}}$ is small.
The difference between the final $A_{\mathrm {PC}}$ and the uncorrected $A^{\mathrm{unc}}_{\mathrm {PC}}$~\cite{n3he20} is only $3\%$; this small improvement in the $A_{\mathrm {PC}}$ does not cause any significant corrections to the published $A_{\mathrm {PV}}= (1.55\pm 0.97~\mathrm {(stat)} \pm 0.24~\mathrm {(sys)})\times 10^{-8}$~\cite{n3he20}. 

This accurate measurement of the $A_{\mathrm {PC}}$ represents an important testing ground for nuclear physics studies, in particular those involving polarized neutrons. We have calculated $A_{\mathrm {PC}}$ using a number of modern nuclear interactions
derived in the framework of chiral EFT. We have found that this observable depends critically on 
the $0^-$ TME. This quantity is in turn very sensitive to the nuclear Hamiltonian, since
the experiment is performed at an energy close to a rather sharp $0^-$ resonance in the $\heq$ spectrum.
Thus, this observable works like a magnifying glass for the nuclear dynamics, and in particular for NN P-wave
and 3N interactions, so that this study can give valuable information on
these small components of the interaction and can be very useful in order to construct more accurate nuclear
potentials. This can have a noticeable impact on other studies where the accurate determination of the nuclear matrix elements is crucial, as, for instance, in the case of the neutrinoless double beta decay. 

Finally, we comment on the impact of this study on the PV observable $A_{\mathrm {PV}}$. As we have seen, the difficulty in the prediction
of $A_{\mathrm {PC}}$ is due to the large variability in the TME ${}^0T^{2,1}_{11,11}$. However, this element does not enter
in the calculation of $A_{\mathrm {PV}}$ (it is suppressed with respect to other terms by a factor $\sim q_2$). Therefore, the theoretical
prediction of $A_{\mathrm {PV}}$ is much less sensitive to the choice of the strong Hamiltonian.

%\appendix
%\section{The $\het(\vec {\mathrm n},{\mathrm p})\tri$ parity-conserving asymmetry}
%\label{SupplementalMaterial}
%\date{\today}

{\bf Appendix: The $\het(\vec {\mathrm n},{\mathrm p})\tri$ parity-conserving asymmetry}
%\maketitle

%\section{Notes}
%PRL source on requirements:
%\begin{itemize}
%    \item \url{https://journals.aps.org/prl/authors/supplemental-materials-journals}
%$    \item \url{https://journals.aps.org/authors/supplemental-material-instructions}
%\end{itemize}
%In short don't repeat from paper, don't use as a source of extra space, don't add new references, and make it material that %is useful only to a minority of readers otherwise it goes in the main paper.

%{\bf Ion and electron collection times in detector chamber and how the charge collection defines the uncertainty in $\beta$.}

Here, we give a more detailed discussion on the ion and the electron collection in the chamber and how it defines the main systematic uncertainty in the inverse neutron wavelength $\overline\mu$. The charge collection process is studied by performing Garfield++ simulations in a chamber model~\cite{Mark2016}. After determining the final systematic uncertainty in $\bar\mu$, we calculate the $\beta=A_{\mathrm {PC}} / \bar\mu$ and its value is finally given in Eq.~(7) of the paper.
%\begin{align}
%	\beta = (-1.97\pm 0.28~\mathrm {(stat)}\pm 0.10~\mathrm {(sys)})\times %10^{-6}\;\mathrm{\AA}\,.
% \end{align}

\begin{figure}[tbp]
  \centering
  \begin{tikzpicture}[scale=.350]
    \input{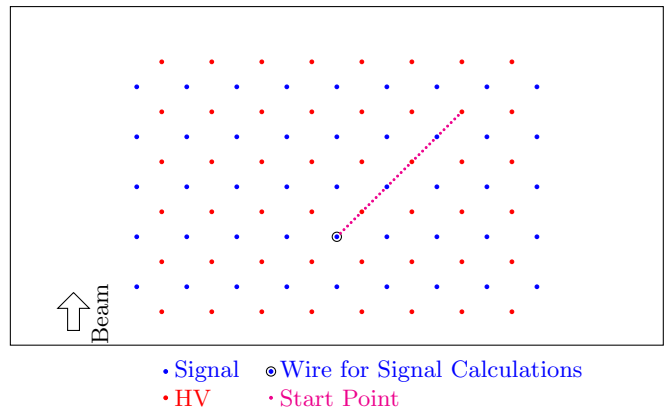}
  \end{tikzpicture}
  \caption{A geometry for the charge collection Garfield++ simulations that is discussed in text. Shown is a scaled cross section of the truncated target chamber model, where signal (blue) and HV (red) wires are perpendicular to the page.  The black outline is the grounded surface corresponding to the inside face of the cylindrical target vessel.  The wire separation and housing are to scale.  The distance between signal (or HV) wire planes is $1.9$~cm. Ion-electron pairs are created on each of the  35 start points of the upwards going diagonal doted line. Garfield++ transports the released charges at $^3$He pressure of $43.6$~kPa and in electrical field produced by $-350$~V on HV wires and calculates induced signals on the indicated wires and average drifting times to reach either a HV or a signal wire. }
  \label{fig:Garf:Layout}
\end{figure}

Fig.~\ref{fig:Garf:Layout} shows a cross sectional view of the truncated chamber model used to study charge collection and cross-talk between wire cells. For charge collection the $^3\mathrm{He}^{+}$-ion and electron pairs are produced on each start point along the diagonal dotted line. Then Garfield++ calculates the time it takes for ions to reach high voltage wires or for electrons to collect onto signal wires~\cite{Mark2016}.  The drift velocities of the ions and the electrons in the chamber are defined by the $^3$He number density and the local electrical field strength produced by -350~V~\cite{Mark2016}. 
%In the $^3$He gas density of the chamber (at $43.6$~kPa) the proton from decay 
%can range up to $14$~cm thus distributing the ionization charges to several wire cells. 

%A schematic of a cross section view of the chamber where the signal (blue) and HV (red) wire arrangement is shown. Distance between signal (or HV) wire planes is $1.9$~cm. A n-$^3$He capture takes place in location of signal wire "a" and then the proton track goes diagonally through 2.5 wire cells to point "e", about a 7~cm long distance.
 
\begin{figure}
	\includegraphics[trim={0 3.2 0 0},clip,width=0.48\textwidth]
        {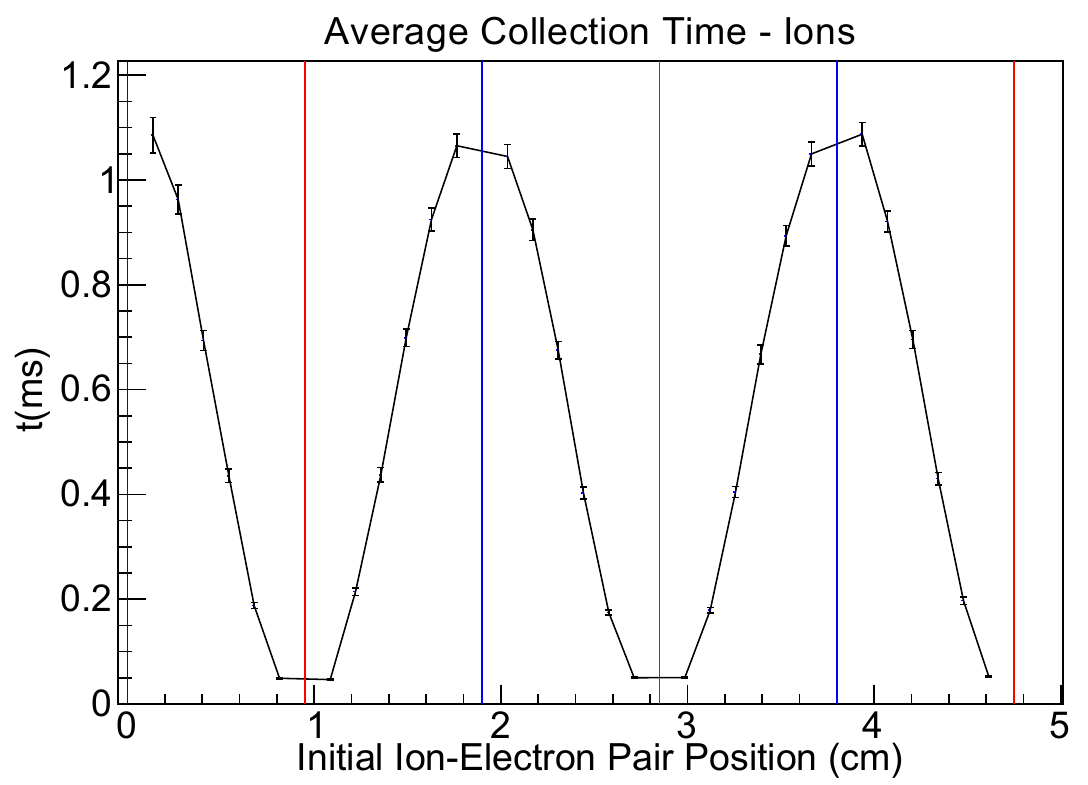}
\caption
{The average $^3\mathrm{He}^{+}$-ion collection times with the geometry of Fig.~\ref{fig:Garf:Layout}. The vertical blue and red lines indicate locations of signal and HV wires.
Collection times are shown for each ion created on 35 dots of the source line with standard deviation. Drift times depend on distance to wire and strength of varying electrical field. Ion-Electron pair position is the horizontal displacement relative to the Wire for Signal Calculations marked in figure \ref{fig:Garf:Layout}. The line connecting the points is just for guiding purpose.}
\label{fig:Garf:IonDiag}
\end{figure}

Fig.~\ref{fig:Garf:IonDiag} shows the calculated average $^3\mathrm{He}^{+}$ collection times with standard deviation in the wire cells as a function of initial location of ion production on the doted line. The created ions are drifting towards HV wires driven by strength of the local electrical fields and collisions with $^3$He atoms in gas. 
%The ions created close to signal wire have the longest distance to travel leading to an %overall longer drift times compared to ions produced closer the HV wires. 
The ion collection times in the simulated geometry can extend up to $1.1$~ms. The Garfield++ results for the electron drifting times have the same shape as ions, but the maxima are on the signal wires and the average collection times are three orders of magnitude shorter than for ions~\cite{Mark2016}. Depending on the location of the start of the ion-electron pair in the chamber, Garfield++ simulations result in the collection times that can be up to 1.8~ms long ~\cite{Mark2016}. When estimating uncertainties in neutron wavelength (or inverse wavelength) calculations, the 1.8~ms ion collection time dominates all the other $TOF$ uncertainties, i.e.\ the electron collection times, the 0.32~ms width of the DAQ time bin and the 0.140~ms FWHM of the emission time distribution from moderator for neutrons with same energy~\cite{emissiontime}.

%\begin{figure}
%	\includegraphics[trim={0 3.2 0 %0},clip,width=0.48\textwidth]
%{Graf-2ndDiag2015_08_10-IonCollTime}
%	\caption{ \textcolor{red}
%{text

%}
%	\label{fig:Garf:IonDiag}
%\end{figure}

%{\color{blue} this is not for text; since electrons are collected very fast, 
%all them are collected, 
%then we should only find what is the ion collection time of when 70 percent of %charge is %collected, 
%which would be about 0.8~ms?}
From distribution of charge collection times, it can be estimated that about $70$~\% of ions are collected at about $0.8$~ms, and since the electron collection times are much shorter, the $0.8$~ms time is used as a charge collection time uncertainty in $TOF$. Adding the $TOF$ uncertainties -- 0.8~ms, 0.32~ms and 0.140~ms -- in quadrature, we obtain that the error due to $TOF$ is $=\pm 0.87$~ms.
%{\color{blue} This note is not meant to go to paper: 
%The lower TOF uncertainty limit is  $  
%\sqrt{(0.14)^2+(0.32)^2}=- 0.35$~ms and the upper limit would be 
%$\sqrt{(0.14)^2+(0.32)^2+0.8^2}= + 
%0.87$~ms. We should use a TOF error of $\pm 0.87$~ms?}
%The lower charge collection time is defined by the widths of the DAQ time bin and 
%the the emission time %uncertainty. Combine errors in quadrature, 
%we have the lower limit as $=-0.35$~ms and the upper limit as %$=0.87$~ms. 

The neutron wavelength is calculated using the relation 
\begin{align}
%mv^2 = \frac{\hbar^2}{m}\frac{1}{\lambda^2}\rightarrow 
\lambda=\frac{\hbar}{mv}=\frac{\hbar {t}}{ml},
\label{eq:WaveLength}
\end{align}
where $m$ is the neutron mass, $l$ is the neutron path length, i.e.\ the distance from the moderator surface to a wire plane in the chamber. This distance is known accurately through surveys. Furthermore, $t$ is the neutron $TOF$, i.e.\ the time it takes the neutron travel to length $l$. As explained in the paper, the $TOF$ measurement starts when the proton beam interacts with the mercury
target. Therefore, the $TOF$ has to be corrected with the neutron emission time ($0.07$~ms, with a weak neutron energy dependence), a time from start of $TOF$ to moment
when the neutron exits the moderator~\cite{emissiontime}. Because of the slow speed of the neutrons and the long distance to detector ($17$~m), the start of the DAQ measurement has to be delayed by $13.500$~ms to sample the wavelength range of interest. The $TOF$ measurement ends when the neutron reaches the wire. The DAQ digitizes the
signals in 49 time bins of 0.32 ms width.

Since the measured $A_{\mathrm {PC}}$ is an average value over the range of wavelengths and $A_{\mathrm {PC}}~\propto~1/\bar\lambda$, then $A_{\mathrm {PC}}(\mu) = \beta \times \bar\mu$, where $\beta$ is a constant that we want to extract from $A_{\mathrm {PC}}$ and $\bar\mu \equiv 1/\bar\lambda$. 

%Next, we determine a reference wavelength and an inverse wavelength for the wavelength %range of the measurement without including uncertainties in distance or in $TOF$. 
%The weighted average wavelength $\bar\lambda$ is calculated for each 16 wire planes %$\mathrm j$ by integrating over 40 time bins $\mathrm i$ in each plane. 
%\begin{align}
%    \bar{\lambda}_{\mathrm j} = \frac{\sum_{\mathrm i} \lambda_{\mathrm {i,j}}I_{\mathrm %
%{i,j}}}{\sum_{\mathrm i} I_{\mathrm i}},
%\end{align}
%where $\lambda_{\mathrm {i,j}}$ is calculated using the distance from the moderator to the %wire plane j and $TOF=13.500-0.07+\mathrm j \times 0.35$~ms, where j is the wire plane %number. $I_{\mathrm {i,j}}$ is the measured detector yield distribution in each time bins %of the plane $\mathrm j$. In each plane has 9 signal wires and $I_{\mathrm {i,j}}$ is a sum %of yields of the wires. 
%To obtain the overall mean wavelength of the chamber, the mean wavelengths of the 16  %planes are combined in a weighted mean, where the weighting factor is over the time bins %integrated yield of the central wire of the plane.

%New version proposed by M. McCrea

Next, we determine the weighted average of the neutron wavelength and the inverse wavelength in the integration time window used for the asymmetry calculation without including uncertainties in the distances or in the $TOF$. 
The weighted average wavelength calculated for each of the 16 wire planes $\mathrm j$ by integrating over 40 time bins $\mathrm i$ in each plane is 
\begin{align}
    \bar{\lambda}^{\mathrm j} = \frac{\sum_{\mathrm i} \lambda^{\mathrm j}_{\mathrm {i}}P^{\mathrm j}_{\mathrm {i}}}{\sum_{\mathrm i} P^{\mathrm j}_{\mathrm i}},
\end{align}
where $\bar\lambda^{\mathrm j}_{\mathrm i}$ is calculated by substituting in Eq.~(1) $l$ with the distance from the moderator surface to the wire plane $\mathrm j$ and $t=TOF=(13.500-0.07)$~ms, the time when DAQ starts sampling. The distance between wire planes is $1.9$~cm. $P^{\mathrm j}_{\mathrm {i}}$ is the measured detector yield distribution in each time bin $\mathrm i$ of the center wire of the plane $\mathrm j$.
To obtain the overall mean wavelength of the chamber, the mean wavelengths of the 16 planes are combined in a weighted mean, where the weighting factor is the yield of the central wire of the plane integrated over the time bin window.

The mean inverse wavelength of the chamber $\bar\mu$, is obtained by replacing variable $\lambda^{\mathrm j}_{\mathrm i}$ in Eq.~(2) with $1/{\lambda^{\mathrm j}_{\mathrm i}}$ and then repeating the process performed for $\bar\lambda$.

For the detector yield weighted mean neutron wavelength and inverse wavelength across 16 wire planes we obtain
\begin{align}
   \bar\lambda&=  4.7775 \pm 0.0317~\mathrm {\AA}~\mathrm {and}\\ 
   \bar{\mu}&=  0.2152 \pm 0.0015~\mathrm {\AA}^{-1}, 
\end{align}
where errors are statistical.
%Seppo, for reference the main error contributions here are 0.1m variation in the mean neutron path length down the guide as simulated by Chris Craford in McStas, and time bin width with distance being dominant.

Finally, we set the upper and lower systematic limits to $\bar\lambda$ and $\bar\mu$, which then determines limits for $\beta$. The $Q$-value of the neutron-$^3$He capture is 765 keV, that is shared as a kinetic energy between the final
state proton, 571 keV,  and the triton, 194 keV.  At 43.6~kPa pressure of $^3$He the proton from the decay can travel up to 14~cm~\cite{Mark2016}. An uncertainty in the neutron path length can be selected to be $\pm 14$~cm, which is close to $1~\%$ uncertainty in the neutron path length, and the $TOF$ uncertainty of $\pm 0.87$~ms is used.
With these systematic uncertainties we calculate the weighted average upper and lower limits to $\bar\lambda$ across the 16 wire planes with the process described above. 
As a result we obtain
\begin{align}
    \bar\lambda=4.78 \pm 0.03~\mathrm{\AA},\\
    \lambda_{\mathrm {max}}=5.01 \pm 0.03~\mathrm{\AA},\\
    \lambda_{\mathrm {min}}=4.74 \pm 0.03~\mathrm{\AA}.
\end{align}
Correspondingly, the weighted average inverse wavelength $\bar\mu$ for the integrated time bins in each wire plane is calculated to be
\begin{align}
    \bar\mu=0.2151 \pm 0.0015~\mathrm{\AA}^{-1},\\
    \mu_{\mathrm {min}}=0.2046 \pm 0.0014~\mathrm{\AA}^{-1},\\
    \mu_{\mathrm {max}}=0.2169 \pm 0.0015~\mathrm{\AA}^{-1}.
\end{align}
With $\bar\mu$ from Eq.(8) and $A_{\mathrm {PC}}$ from Eq. (3) of the paper, we can extract the central value for $\beta$,
\begin{align}
    \beta &= A_{\mathrm{PC}}/\bar{\mu},\\
	 &=  [ -1.972 \pm 0.28\mathrm {(stat)} \pm 0.017\mathrm {(sys)}]\times 10^{-6}\;\mathrm{\AA}^{-1},
\end{align}
while for the upper and lower limits in $\beta$ we get
\begin{align}
  \beta_{\mathrm {min}} &=  [-0.2074  +/- 0.0279] \times 10^{-6}~\mathrm{\AA}^{-1},\\
  \beta_{\mathrm {max}} &=  [-0.1956  +/- 0.0279] \times 10^{-6}~\mathrm{\AA}^{-1}.
\end{align}
The difference between the minimal and maximal values of $\beta$ is then used to set a stronger systematic 
uncertainty in $\beta$,
\begin{align}
	 \Delta\beta=& |\beta_{\mathrm {max}}-\beta_{\mathrm {min}}|\\
             =&  0.1179\times 10^{-6}\;\mathrm{\AA}^{-1}.
\end{align}
Combining the systematic uncertainties of Eqs~(12) and~(16) in quadrature, we obtain the final uncertainty for $\beta$ of $0.119 \times 10^{-6}$.
This gives the final $\beta$-value of
\begin{align}
	\beta =& [-1.97 \pm 0.28\mathrm {(stat)} \pm 0.12\mathrm {(sys)}]\times 10^{-6}\;\mathrm{\AA}^{-1}.
\end{align}

{\bf Acknowledgements} 
We gratefully acknowledge the support of the U.S. Department of Energy Office of Nuclear Physics through grant
No. DE-FG02-03ER41258, DE-AC05-00OR22725, DE-SC0008107 and DE-SC0014622, the US National Science Foundation award No: PHY-0855584,
the Natural Sciences and Engineering Research Council of Canada (NSERC), the Canadian Foundation for Innovation (CFI), and the Mexican PAPIIT-UNAM award No. IN11913 and AG102023.
This research used resources of the SNS of ORNL, a DOE Office of Science User Facility.
% ADDITIONAL INPUTS HERE

\end{document}